# Comparison of Single Carrier FTN-QAM and PCS-QAM for Amplifier-less Coherent Communication Systems

Dongdong Zou[1,*], Fan Li[2], Wei Wang[2], Zhongxing Tian[1], Yuheng Liu[2], Gangxiang Shen[1], and Yi Cai[1]

1)The School of Electronic and Information Engineering, Soochow University, Suzhou 215006, China
2)The School of Electronics and Information Technology, Guangdong Provincial Key Laboratory of Optoelectronic Information Processing Chips and Systems, Sun Yat-Sen University, Guangzhou 510275, China
Email: ddzou@suda.edu.cn

**Abstract:** A performance comparison of FTN-QAM and PCS-QAM for amplifier-less short-reach coherent communication systems is provided. With the applications of phase tracking partial response DFE and turbo equalization strategy, FTN-16QAM exhibits about 0.9dB power margin advantage over PCS-64QAM.

## 1. Introduction

The proliferation of diverse artificial intelligence (AI) applications is driving an exponential increase in data traffic, rendering the continuous enhancement of transmission capacity as a fundamental necessity within optical communication systems. This necessity is particularly pronounced in short-reach fiber optical links, notably next-generation data center interconnects (DCI), which demand ultra-high data rates (e.g., 3.2Tb/s and beyond) while operating under severe power consumption and physical space constraints. The coherent communication technique has been considered as a highly competitive solution for next generation ultra-high speed DCIs, owing to its intrinsic capability to exploit higher degrees of freedom in data delivery. Combined with advanced high spectral efficiency (SE) signaling schemes, such as probabilistic constellation shaping (PCS) [1] and faster-than-Nyquist (FTN) modulation [2], coherent optical communication has been widely explored in short-reach DCIs [3].

High-order modulation format signaling coupled with the PCS technique has been extensively demonstrated in long-haul coherent optical communication systems [4]. However, for short-reach DCIs, the optical amplifier is not equipped due to the low transmission loss and system cost limitation. The proper implementation of PCS within the amplifier-less coherent communication system warrants further in-depth investigation, even though significant works utilizing the Maxwell-Boltzmann (MB) distribution have already demonstrated performance gains [5]. As another high SE signaling scheme, FTN accelerates the data transmission by deliberately introducing inter-symbol-interference (ISI) to break the Nyquist criterion while without compromising the system's minimum Euclidean distance [6]. However, the intentionally introduced inter-symbol interference (ISI) renders conventional coherent digital signal processing (DSP) schemes ineffective. Specifically, the absence of a distinct constellation contour in the blindly equalized signal precludes the reliable operation of both the frequency offset estimation (FOE) and carrier phase estimation (CPE). Therefore, a highly effective coherent DSP architecture is required in coherent FTN systems, especially with high compression factor.

In this paper, a novel high-performance FTN signaling scheme is proposed for amplifier-less short-reach coherent communication systems, which is enabled by a phase tracking partial response decision feedback equalizer (PT-PRDFE) and modified turbo equalization. Moreover, a fair performance comparison between PCS-64QAM (H=5) and FTN-16QAM ($α=0.8$) is carried out. According to the experimental results, the proposed FTN-16QAM signaling scheme exhibits about 0.9dB power margin improvement compared to the MB PCS-64QAM over 40km amplifier-less single-mode fiber (SMF) transmission.

## 2. Principle

In order to achieve the high-performance FTN signaling, a novel DSP architecture tailored for single-carrier FTN-QAM systems is proposed, including pilot-tone-enabled FOE, PT-PRDFE, and modified turbo equalization for further performance enhancement. Fig. 1 shows the scheme diagram of the proposed PT-PRDFE, which is applied to simultaneously perform the equalization of the FTN-16QAM signal to a partial response 16QAM signal and the compensation of phase noise. Compared with the full response equalization rule, partial response equalization can achieve a balance between amplified noise and eliminated ISI. The detailed information about the phase tracking equalizer can be found in [7], in which the phase tracking feedforward equalizer is demonstrated.

To further improve the performance of the FTN system, a modified turbo equalization is cascaded following the PT-PRDFE. As shown in Fig. 2, a poster-filter is applied to whiten the residual colored noise. The signal is then fed into the turbo equalization module. Generally, the trellis diagram of the sequence detector should be established based on the partial response 16QAM signal (or PAM4 for I/Q branch) and the post-filter. However, achieving reliable soft information exchange between the sequence detector and the FEC decoder is highly challenging in this way, since the former one operates on the partial response signal while the latter processes the original bit stream, corresponding to original 16QAM signal, directly. In this paper, the trellis diagram of the sequence detector is established according to the original 16QAM signal and the final effective filter $h_{SD}$, $h_{SD} = h_{PR} * h_{PF}$. In this way, the input and output information of both the sequence detector and the FEC decoder pertain to the original 16QAM symbols, which facilitates a convenient implementation of the iterative soft information exchange. Moreover, as the complexity of the sequence detector increases exponentially with the modulation format, meaning that the proposed scheme shows a much lower complexity.

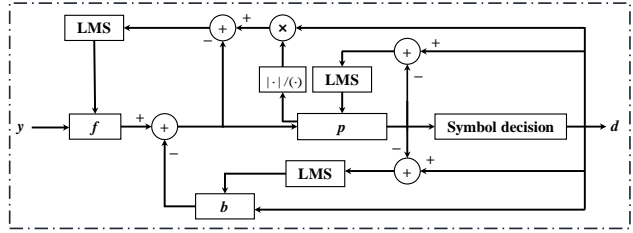

Fig. 1. Scheme diagram of phase tracking partial response DFE.

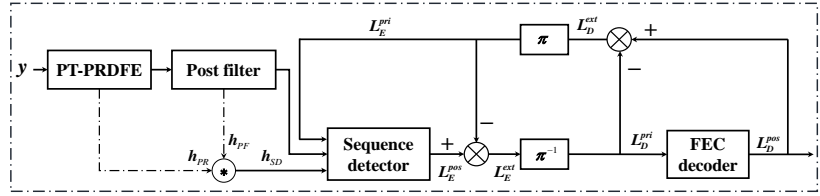

Fig. 2. Modified turbo equalization scheme for the single-carrier FTN-QAM system.

## 3. Experimental Setup and Results

A detailed performance comparison of two kinds of high SE signaling schemes is demonstrated in 45Gbaud FTN-16QAM with a compression factor of 0.8 and 36Gbaud PCS-64QAM with an entropy of 5, in which their SE and occupied bandwidth are identical, facilitating a fair performance comparison. For PCS-64QAM, MB and IvMB distributions are discussed. The probabilistic distributions of MB and IvMB PCS-64QAM are illustrated in insets (*i*) and (*ii*) of Fig. 3. The experimental setup and the detailed DSP flow are shown in Fig. 3. At the transmitter, the signal is generated by an arbitrary waveform generator (AWG) operated at 64GSa/s (Keysight M8195). The electrical signals output from the AWG are then loaded to a dual-polarization IQ modulator (DP-IQM) to realize the electro-optical conversion, and the optical carrier is generated from an external cavity laser (ECL) with a linewidth of 100kHz and 16dBm optical power. As the purpose of this paper is to evaluate the performance of the two signaling schemes in short-reach amplifier-less coherent systems, there is no optical amplifier utilized in transceivers. At the receiver, another ECL is used as the local oscillator laser (LO) with an output power of 16dBm. Then, the incoming signal and the LO are input into an integrated coherent receiver (ICR) for coherent detection. The detected signals are captured by the Fujitsu analog-to-digital converter (ADC) with a sampling rate of 80GSa/s. Both the Tx-DSP and Rx-DSP are achieved offline in MATLAB. In the Tx-DSP, the pseudo-random binary sequence (PRBS) is processed by probabilistic amplitude shaping (PAS) or the FEC encoder and QAM mapper. Subsequently, a pulse shaping filter is applied, followed by a frequency pilot-tone inserting for frequency offset estimation. The generated PCS-64QAM or FTN-16QAM signal is resampled to match the sampling rate of the AWG. Inset (*iii*) of Fig.3 depicts the electrical spectrum of the typical 45Gbaud 16QAM and FTN-16QAM ($\alpha$=0.8) signals. A significant bandwidth compression can be observed. At the receiver side, the DSPs include resampling, FOE, chromatic dispersion compensation (CDC), retiming, synchronization, phase tracking equalization, PAS decoding or turbo equalization. Finally, the bit error rate (BER) is counted. In the FTN-16QAM system, a $2^{nd}$ order PT-PRDFE is applied, and the constellation of $2^{nd}$ PR 16QAM signal is depicted in the inset (*iv*) of Fig. 3. A nonuniform 49-QAM signal is observed.

In an amplifier-less coherent communication system, the system is primarily limited by the launched or received optical power (ROP) rather than the optical signal-to-noise ratio (OSNR). Consequently, the ROP serves as the appropriate channel performance metric. According to the experimental results, the launched optical power of 36Gbaud MB PCS-64QAM, 36Gbaud IvMB PCS-64QAM, and 45GBaud FTN 16QAM signals are −14dBm, −11dBm, and −11.8dBm, respectively. To ensure a fair comparison, the BER versus the power margin (defined as the difference between the maximum ROP and the actual ROP) of different signals under optical back-to-back (OBTB) transmission is evaluated. Although the IvMB PCS-64QAM signal exhibits the highest launched power, it shows the worst BER performance, as illustrated in Fig. 4(a). The FTN-16QAM signal shows a comparable performance to the MB PCS-64QAM signal in the low power margin region. However, it demonstrates a certain performance penalty at high power margins when only the BCJR sequence detector, rather than turbo equalization, is employed. Then, the BER versus power margin of different signals over 40km SMF transmission is depicted in Fig. 4(b). The FTN-16QAM shows about 1.2dB power margin penalty compared to the MB PCS-64QAM signal. For FTN signaling, turbo equalization has been widely discussed to enhance the system performance. As presented in Section II, a novel modified turbo equalization scheme is also proposed in this paper. Fig. 4(c) shows the BER of FTN-16QAM over 40km SMF transmission versus the iterative number in turbo equalization. At a 4dB power margin, 4 iterations are enough to achieve error-free transmission. Finally, the BER performance of PCS-64QAM and FTN-16QAM with turbo equalization is demonstrated in Fig. 4(d). With the proposed modified turbo equalization scheme, FTN-16QAM shows about 0.9dB power margin advantage compared to the MB PCS-64QAM.

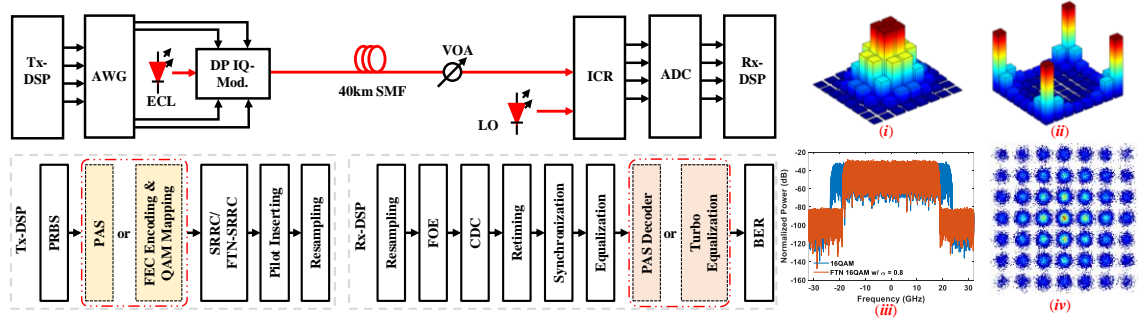

Fig. 3. Experimental setup and transceiver DSPs. Probability of (i) MB PCS-64QAM and (ii) IvMB PCS-64QAM with entropy of 5; (iii) Electrical spectrum of typical 16QAM and FTN-16QAM ($\alpha$=0.8) signals; (iv) Constellation of $2^{nd}$ order partial response 16QAM signal.

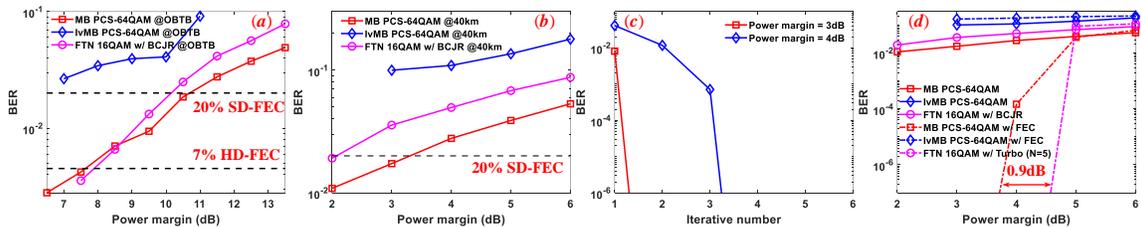

Fig. 4. BER versus power margin of PCS-64QAM and FTN-16QAM (a) at OBTB, (b) over 40km SMF transmission; (c) BER versus turbo iterative number of FTN-16QAM over 40km SMF transmission; (d) BER versus power margin of PCS-64QAM and FTN-16QAM with FEC or turbo equalization.

## 4. Conclusion

In this work, a novel high-performance FTN signaling scheme is proposed for amplifier-less short-reach coherent communication systems, which is enabled by a PT-PRDFE and modified turbo equalization. According to the results, the proposed FTN-16QAM signaling scheme exhibits a 0.9dB power margin improvement relative to the MB PCS-64QAM. This superior performance firmly validates the proposed technique as a promising solution for future high SE short-reach optical interconnects.

## 5. Acknowledge

This work is partly supported by the National Key Research and Development Program of China (2023YFB2906000); National Natural Science Foundation of China (62501414, 62275185, 62271517); Jiangsu Funding Program for Excellent Postdoctoral Talent (2025ZB335).